\documentclass[aoas,MSNbibl,nameyear,seceqn,dvips]{arximspdf}
\usepackage{graphicx}

\doi{10.1214/10-AOAS389}
\volume{5}
\issue{1}
\pubyear{2011}
\firstpage{364}
\lastpage{380}

\makeatletter
\newcommand{\eqref}[1]{(\ref{#1})}
\makeatother

\begin{document}
\begin{frontmatter}

\title{Latent rank change detection for analysis of splice-junction
microarrays with nonlinear effects}
\runtitle{Latent rank change detection}

\begin{aug}
\author[A]{\fnms{Jonathan} \snm{Gelfond}\corref{}\ead[label=e1]{gelfondjal@uthscsa.edu}},
\author[A]{\fnms{Lee Ann} \snm{Zarzabal}\ead[label=e2]{zarzabal@uthscsa.edu}},
\author[A]{\fnms{Tarea} \snm{Burton}\ead[label=e3]{burtont@uthscsa.edu}},
\author[A]{\fnms{Suzanne} \snm{Burns}\ead[label=e4]{burnss@uthscsa.edu}},
\author[B]{\fnms{Mari} \snm{Sogayar}\ead[label=e5]{Mcsoga@iq.usp.br}}
\and
\author[A]{\fnms{Luiz O. F.} \snm{Penalva}\ead[label=e6]{penalva@uthscsa.edu}\thanksref{t1}} %

\thankstext{t1}{Corresponding author.}
\runauthor{J. Gelfond et al.}

\affiliation{UT Health Science Center San Antonio,
UT Health Science Center San Antonio,
UT Health Science Center San Antonio,
UT Health Science Center San Antonio,
Universidad de S\~ao Paulo and
UT Health Science Center San Antonio}

\address[A]{J. Gelfond\\
L. A. Zarzabal\\
T. Burton\\
S. Burns\\
L. O. F. Penalva\\
UT Health Science Center San Antonio\\
Mail Code 7933\\
7703 Floyd Curl Drive\\
San Antonio, Texas 78229-3900\\
USA\\
\printead{e1}\\
\phantom{E-mail: }\printead*{e2}\\
\phantom{E-mail: }\printead*{e3}\\
\phantom{E-mail: }\printead*{e4}\\
\phantom{E-mail: }\printead*{e6}}

\address[B]{M. Sogayar\\
Universidad de S\~ao Paulo\\
Av. Prof. Lineu Prestes\\
748 - Butant\~{a}\\
Caixa Postal 26077\\
CEP 05513-970, S\~{a}o Paulo, SP\\
Brazil\\
\printead{e5}}

\end{aug}

\received{\smonth{12} \syear{2009}}
\revised{\smonth{6} \syear{2010}}

%
\begin{abstract}
Alternative splicing of gene transcripts greatly expands the functional
capacity of the genome, and certain splice isoforms may indicate specific
disease states such as cancer.
Splice junction microarrays interrogate thousands of splice junctions, but
data analysis is difficult and error prone because of the increased
complexity compared to differential gene expression analysis.
We present Rank Change Detection (RCD) as a method to identify
differential splicing events based upon a straightforward probabilistic
model comparing the over- or underrepresentation
of two or more competing isoforms. RCD has advantages over commonly
used methods because it is robust to false positive errors due to
nonlinear trends in microarray measurements.
Further, RCD does not depend on prior knowledge of splice isoforms, yet
it takes advantage of the inherent structure of mutually exclusive
junctions, and
it is conceptually generalizable to other types of splicing arrays or RNA-Seq.
RCD specifically identifies the biologically important cases when a
splice junction becomes more or less prevalent compared to other
mutually exclusive junctions.
The example data is from different cell lines of glioblastoma tumors
assayed with Agilent microarrays.
\end{abstract}

%
\begin{keyword}
\kwd{Alternative splicing}
\kwd{gene expression analysis}
\kwd{microarray}.
\end{keyword}

\end{frontmatter}

\section{Introduction}

\begin{figure}

\includegraphics{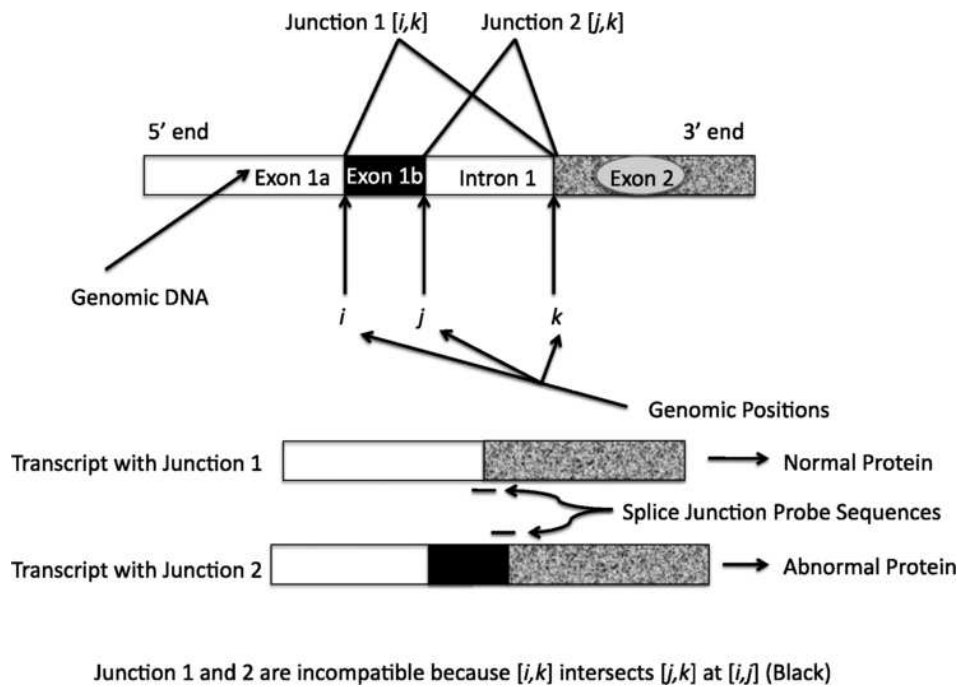}

\caption{Schematic of splicing process, incompatible junctions and
junction probes.
The gene is shown at the top. The splicing process has two
possibilities. First,
remove exon 1b (Black) and intron~1. The fusion between exons 1a and~2
results in Junction 1. Second, only remove intron 1 while
retaining exon 1b. The fusion between exons 1b and 2
results in Junction 2. The retention of exon 1b is translated into an
abnormal protein,
and higher fluorescence of the Junction 2 probe relative to the
Junction 1 probe indicates this abnormality.}\label{fig:schematic}
\end{figure}

Genomic DNA contains the sequential codes for constructing proteins
that perform biochemical and structural tasks essential for life. The
DNA of a gene is first transcribed into a pre-messenger RNA (pre-mRNA)
transcript. The rate of transcription is referred to as gene expression
and is a measure of the gene's level of activity. The DNA fragment
encoding a particular gene does not get entirely converted into mRNA. A
gene is composed of sequence blocks called introns and exons. The
introns are removed or spliced out of the pre-mRNA sequence, whereas
the exons are retained to form the mature mRNA which is later
translated into proteins. However, some exons are selectively included
in the mature mRNA so that different splicing variants or isoforms
result. The different splicing variants can give rise to different
protein isoforms which sometimes display different structural and/or
functional properties [\citet{Stryer95}]; see Figure~\ref{fig:schematic}.
This is known as \textit{alternative splicing}. There are
several classes of alternative splicing events. For example, exon
skipping occurs when an exon is selectively ``skipped'' or excluded in
the mature mRNA. See Supplementary Figure 1 for an illustration of some
common splicing events.

In the late 80's and early 90's, only a few alternative splicing events
were known.
Alternative splicing was
considered a rare event, and its importance misunderstood. Current
estimates of the number of human genes that
have at least 2 splice variants range from 60 to 85\% [reviewed in
\citet{Cuperlovicetal06}].
Different from other types of gene regulation, alternative splicing
does more than simply modulate
the levels of expression of affected genes.
In extreme cases like the
Dscam gene in Drosophila, the number of potential splice variants is
equivalent to almost double
the total number of genes present in the Drosophila genome [\citet
{CelottoGraveley01}].
Several different splicing events including exon skipping events,
mutually exclusive exons and alternative $5'$ and $3'$ splice sites
form the repertoire of alternative splicing.
About 70--88\% of differential splicing events affect the coding region,
and the numbers of known splicing events
will increase thanks to technologies such as alternative splicing
arrays and RNA-Seq [\citet{Sultan2008}].

Splicing arrays contain tens or hundreds of thousands of
probes that are complementary to exons or splice junctions (the fusion
between exons) of the different splice isoforms of many genes.
These probes bind specifically to segments within a transcript and
quantify the concentrations of specific exons and splice junctions.
Splicing arrays provide higher resolution measurements across the
length of the transcript compared to standard expression arrays that
bind to
a smaller, selective portion of the transcript.
See Figure \ref{fig:schematic} for the placement of splice junction probes.
For a review of the different classes of alternative splicing
microarrays see \citet{moore2008}.
The analysis of alternative splicing arrays is an area in need of
improvement due to difficulties in estimating the relative
concentrations of isoforms with or without
a fully known sequence.
Currently, there are several statistical methods available, but none is
satisfactory because of limitations in application and inaccuracy [see
\citet{Cuperlovicetal06}].
The basic goal of the methods is to estimate the proportions of splice
variants within a tissue and to compare these proportions between tissues.
For example, we would like to estimate the relative prevalence of two
isoforms in one tissue (say, a 1:2 ratio) and compare them to the
relative prevalence in another
tissue (say, a 2:1 ratio). If these relative prevalences are different,
then we have an occurrence of differential splicing between the tissues.
The analysis of alternative splicing events is a more complicated task
than quantifying the overall level of expression of a gene as typically
done by expression arrays.
In expression analyses that compare the overall expression of a given
gene in two tissues (e.g., cancer and normal), three scenarios are
possible: higher expression in cancer, higher expression in normal, and
equal expression. These measurements can be obtained, in principle,
with one probe per gene. In alternative splicing arrays, there is a set
of probes for each alternative splicing event; they normally cover
splice junctions and/or the region that is alternatively spliced. A
substantial percentage of human genes have two or more alternative
splicing events that need to be evaluated individually. The resultant
isoforms are multivariate components of the gene's expression.
Each of these isoforms itself may be up- (or down) regulated in a
comparison between tissues.
Further, if all of the isoforms are up- (or down) regulated, then one
may additionally ask whether they
occur in the same proportion in different tissues. Again, the comparisons
of the relative proportions of isoforms between tissues is one of the
central issues in alternative splicing analysis.\looseness=1


The increase in transcript information is the motivation for examining
alternative splicing in cancerous tissue,
but the added statistical complexity of the data requires novel and
robust analytical methods.
\citet{Srinivasan2005} developed the Splicing Index statistic for exon
arrays. This method estimates
the proportion of alternative isoforms by assuming a linear
relationship between mRNA concentration and probe intensity.
The analysis of splice variation (ANOSVA) proposed by \citet{Cline2005}
is a similar method that uses a parametric model.
ANOSVA uses a two-way analysis of variance model with an interaction where
the first factor is the tissue type level (tumor, normal, etc.) and the
second factor is the probe (splice junction 1, splice junction 2, etc.).
In ANOSVA, the interaction between treatment effect and the probe
effect corresponds to a differential splicing event (DSE).
However, like the Splicing Index, ANOSVA depends on the linearity of
the intensity response curve and
may have a high false positive detection rate. \citet{Shai2006}
presented GenASAP as another analysis method, but
this potentially powerful tool has limited applicability.
GenASAP uses a latent variable Bayesian model and machine learning
techniques to estimate the percentage of
each splice variant in a sample. Although GenASAP was designed to
distinguish between only two isoforms,
many genes may have multiple isoforms, and this limits the GenASAP
method to a fraction of genes that
only have two isoforms or to those genes that are determined by other
methods to have only two prominent isoforms.
\citet{Xing2008} have created normalization and analysis methods for
exon array data. While
their proposed normalization technique applies to splice junction
arrays, the splice junction array
data will likely benefit from models specifically designed for
junctions. Splice junction arrays differ from
exon arrays in that the probe sequences correspond to the junctions
between fused exons in
the final transcript. The number of possible splice junctions exceeds
the number of exons
because if the number of exons is $E$, the number of junctions is
potentially ${E \choose 2}$, not counting partial exons, but
there is a lack of knowledge about which junctions actually occur in
nature. Further, unlike exons,
certain splice junctions are incompatible within the same transcript
based upon the
topology of splicing because splicing
preserves the order of the exons. For example, a transcript with a
junction fusing exons 2 and 4 is
incompatible with a junction fusing exons 2 and 3 because
exon 3 is excluded by junctions 2--4.
Our proposed method is designed to accommodate these features
of junction arrays.

We present Rank Change Detection (RCD) as a method for identifying
differential splicing events (DSEs) based upon a Bayesian model
comparing the over or underrepresentation of two or more competing
isoforms. RCD has advantages over commonly used methods
as it is robust to false positive errors due to nonlinear trends in
microarray measurements and it tests hypotheses of inherent biological interest.
\citet{Gaidatzis2009} have recently shown that these nonlinear effects
in Affymetrix exon arrays are a source of inaccuracy using current
linear models, and we
show that such nonlinearity is found in two color splice junction
arrays as well.
RCD does not depend on prior knowledge of splicing isoforms, yet it
takes advantage of the inherent structure of mutually exclusive junctions.
This method may easily be adapted to multiple platforms including other
types of alternative splicing microarrays or RNA-Seq data.

\section{Experimental design and data structure}\label{sec.def1}

The objective is to identify splicing differences between normal and
tumor tissues or cell lines. In our experiment the normal cell line is
the glial cell
line known as FGG, which is compared on each array to one of four
glioblastoma tumor types.
This particular experimental design is called a \textit{reference design}
because the normal sample is used as a comparator on
each array [\citet{kerr01}].
The two-color microarrays have probes that bind to splice junctions
simultaneously within two samples that are labeled to fluoresce at
green and red wavelengths using Cyanine-5 (Cy5) dye and Cyanine-3 (Cy3) dye,
respectively.
The intensity of fluorescence quantifies the
amount of RNA expression. The dye orientation is balanced so that for
each tumor type, the normal and tumor samples are
labeled with Cy3 and Cy5, respectively, or Cy5 and Cy3, respectively,
in an equal number of replicates.
This is called a \textit{dye-swap} design, and the experimental layout is
shown in Supplemental Table 1.
The log-probe intensity data for a given gene are denoted as
$y_{tjrds}$ where the indices $t$, $j$, $r$, $d$ and $s$ represent tissue,
junction, replicate, dye and spot respectively. Note that $t$ refers to
the type of cells sampled whether from a tissue or a cell line or
culture, but
we will refer to $t$ as tissue without loss of generality.
We found that there was little to no dye bias so we drop this
distinction (index $d$) for convenience.
The spot or probe effect induces correlation between
the red and the green channels that are measured for each probe.
The junction $j$ is defined by the location (in DNA base pairs or bp)
from the transcription start site (or another reference site)
of the $5'$ and the $3'$ sites ($j_5$ and $j_3$) that define the \textit{
interval} $[j_5,j_3]$ of the excised segment of the transcript.
Two junctions $j$ and $j'$ are mutually \textit{incompatible} if
$[j_5,j_3] \cap[j'_5,j'_3] \ne\varnothing$. The nonempty
intersection indicates that one junction excises a component of the
other, thus making mutually exclusive
within the same transcript; see Figure \ref{fig:schematic}. The set of
all junctions incompatible with junction $j$ is denoted as $O_j$.
This implies that the junction $j$ is in the set of incompatible
junctions $O_j$ ($j \in O_j$).
There are possibly more incompatible
junctions in the set $O_j$ if one had knowledge of biologically
plausible transcripts and their corresponding junctions.
Our method of defining incompatible sets
has the advantage that it does not require that the full transcript
sequences are known. The specification of these sets
is critical because this defines the set of junctions whose proportions
are being compared. One typically compares the
relative prevalence $p_{tj'}$
of the junctions $j' \in O_j$ for each tissue $t_1,   t_2$
where $\sum_{j' \in O_j} p_{tj'} = 1$, and the null hypothesis
for a differential splicing event (DSE) is $p_{t_1j'}=p_{t_2j'}$ for
$j' \in O_j$.

\section{Rank change detection method}\label{sec.def}

\begin{figure}

\includegraphics{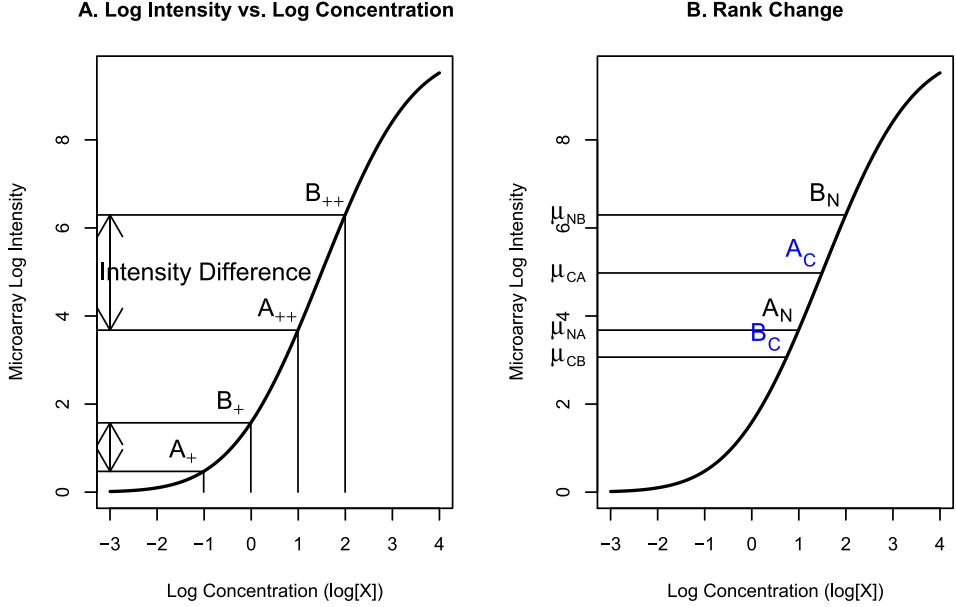}

\caption{(\textup{A}) Example of sigmoidal response of microarray Intensity:
Two incompatible isoforms A and B present in 1:2 a ratio at high ($++$)
and low ($+$) levels of
overall gene expression. Notice that the intensity difference between
the two
isoforms is narrowed considerably in lower concentration
despite the constant concentration ratio of 2:1.
Models assuming linearity could falsely estimate that the ratio of isoforms
had narrowed from 2:1 to closer to 1:1.
(\textup{B}) Rank Changes of isoform intensity are invariant under monotonic
transformation.
There are two incompatible isoforms A and B present in normal (N),
cancer (C). Isoform A is more prevalent in cancer, while isoform B is more
prevalent in normal. The proposed method identifies such changes in
prevalence rankings of isoforms. The mean intensities are shown on the
vertical axis as $\mu_{tj}$.}
\label{fig:sigmaplot5}
\end{figure}

\begin{figure}

\includegraphics{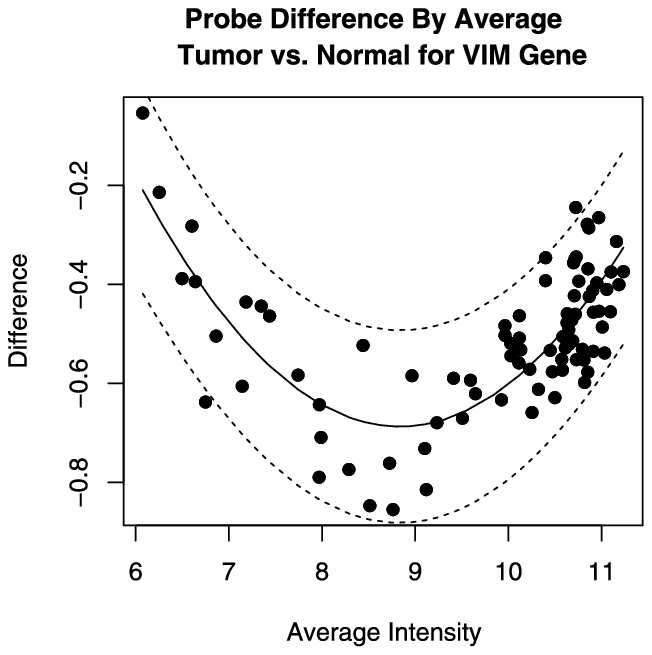}

\caption{Difference vs Average plot for splice junctions of theVIM gene.
Each point is the average value of a specific junction.
The horizontal axis is the estimate of the mean junction intensity
$\frac{1}{2} (\hat{\mu}_{Cj}+\hat{\mu}_{Nj})$, and
the vertical axis is the estimate of the mean junction differences
($\hat{\mu}_{Cj}-\hat{\mu}_{Nj}$).
Note the substantial nonlinear effect apparent in the parabolic trend.
This invalidates
algorithms based upon linearity such as ANSOVA likely resulting in
false positives.}\label{fig:MvA}
\end{figure}

The model of \citet{Cline2005} is an ANOVA model
described by $E[y_{tjr} ] =\mu_{tj}= \mu_0 + \alpha_{t} + \beta_{j} +
\gamma_{t \times j}$ and $\alpha$, $\beta$ and $\gamma$
are the effects of tissue, junction and the interaction between the
tissue and junction.
The rejection of the test of $\gamma_{t \times j }= 0$ implies a
differential splicing
event. That is, the $\gamma_{t \times j }$ represents the relative
increase or decrease
in the junction $j$ prevalence in tissue $t$.
However, the model becomes
invalid in the presence of nonlinearity.
For example, if there is not a DSE and there are two incompatible
junctions ($A$~and $B$), then we denote the mean intensities of the
probes by
$\mu_{tj,}$ say, $\mu_{NA}$, $\mu_{NB}$, $\mu_{CA}$ and $\mu_{CB}$
where $N$ and $C$ are the normal and cancer tissue indices.
According to the ANOSVA model, $\mu_{tj} = \mu_0 + \alpha_{t} + \beta
_{j} + \gamma_{t \times j}$, so that if
$ \gamma_{t \times j}=0$, then $\mu_{CA}-\mu_{NA}=\mu_{CB}-\mu_{NB}$.
Equivalently, if there is no DSE, then
the difference between junction probe mean intensities should be equal.
Substantial nonlinearity results in an extremely high false positive
rate due to the phenomenon in Figure \ref{fig:sigmaplot5}A.
For example, we examine our data for the VIM gene shown in
Figure \ref{fig:MvA}. On the vertical axis is the estimate of the mean
junction differences ($\hat{\mu}_{Cj}-\hat{\mu}_{Nj}$),
and on the horizontal axis is the estimate of the mean junction
intensity $\frac{1}{2} (\hat{\mu}_{Cj}+\hat{\mu}_{Nj})$.
Clearly, the difference $\hat{\mu}_{Cj}-\hat{\mu}_{Nj}$ is not equal
for all junctions with the high and low intensity
junctions having differences closer to 0. This dependence on intensity
is consistent with nonlinear response as in Figure \ref{fig:sigmaplot5}A.
The ANOSVA test for $\gamma_{t \times j}=0$ results in a $p$-value
$<10^{-19}$, which would likely yield a false positive result because
nonlinear response to differential expression is confounded with the
hypotheses about junction by tissue interactions.

We propose a model for rank change that tests the hypothesis that a junction
has the same rank in relative proportion in one tissue compared to
another, as measured by probe intensity.
We define the rank of $p_{tj}$ as $R(p_{tj})=\sum_{j' \in O_j}
I[p_{tj'} \le p_{tj}]$
where $I[\cdot]$ is the indicator function.
Instead of the null hypothesis that proportions themselves are equal,
$p_{t_1j'}=p_{t_2j'}$ for $j' \in O_j$, we
will test that the ranks of the $p_{tj}$ are equal,
$R(p_{t_1j'})=R(p_{t_2j'})$ for $j' \in O_j$, by making
the reasonable assumption that the ranks in mean junctions intensities
$\mu_{tj}$
approximate the ranks of the junction proportions $p_{tj}$.
A shift in rank represents
a decrease or increase in the prevalence of the isoform as shown in
Figure \ref{fig:sigmaplot5}B.
The justification for assessing ranks is that ranks are preserved under
monotonic transformation
due to nonlinear response,
whereas
differences in intensity (i.e., $\gamma_{C \times j}=\hat{\mu}_{Cj}-\hat
{\mu}_{Nj}$) are not.
We propose the following random effects implementation of the ANOSVA
model for two color arrays:
%
\begin{equation}
y_{tjrs} = \mu_{tj} + \nu_{tjrs} +\varepsilon_{tjrs},
\end{equation}
where the $\mu_{tj}$ are the means of the $j$th junction in $t$th
tissue and $\varepsilon_{tjr}$ is an independent Gaussian noise error
term. The term $\nu_{tjrs}$ is a Gaussian random effect corresponding
to a probe on an array, and this
random effect is two-dimensional corresponding to the red and green dyes.
The main idea of detecting rank change comes from the following
example. If there is not a DSE and there are only two junctions ($A$ and
$B$), then the ranks of the mean intensities ($
\mu_{tj'},   j' \in O_j$) are preserved.
In other words, $\mu_{NA}<\mu_{NB}$ implies that $\mu_{CA}<\mu_{CB}$
where $N$ and $C$ are the normal and cancer tissue indices. This is
equivalent to saying junction $B$ is more prevalent than junction $A$ in
both tissues, assuming that intensities of the two junction probes
roughly reflect their relative concentrations.
If the ranks of the junction prevalences change as in Figure \ref
{fig:sigmaplot5}, then there is a DSE.
For an arbitrary number of isoforms, no DSE implies that the ranks of
the $\mu_{tj}$ are
preserved across tissues. We do not directly observe the ranks of the
latent means $\mu_{tj}$ as there is
variability in measurement, but we can estimate the latent ranks of the
means from the posterior distribution.
Let the rank of the mean intensity $\mu_{tj}$ of junction $j$ within a
tissue $t$ be defined
as $R(\mu_{tj})=\sum_{j' \in O_j} I[\mu_{tj'} \le\mu_{tj}]$. Our null
hypothesis for a DSE
between tissues $t_1$ and $t_2$ would be that
\[
R(\mu_{t_1j})=R(\mu_{t_2j} ).
\]
That is, the rank of the average intensity of junction $j$ relative to
other incompatible junctions $j' \in O_j$ is preserved across tissues.
Since there are 40,000 models corresponding to each junction on the
chip, we will estimate the posterior distributions using approximations
to accelerate model fitting.
We can approximate the posterior distributions of $\mu_{tj}$ by the
maximum likelihood theory point and variance estimates ($\hat{\mu
}_{tj}$ and $\hat{\Sigma}_{\mu}$).
We estimate and compare the posterior distributions of $R(\mu_{t_1j})$
and $R(\mu_{t_2j})$
with Monte-Carlo integration of the posterior distribution
$p([\mu_{t1},
\ldots,\mu_{tJ}]|\mathrm{Data})$ where $J=$ size of $O_j$.
The posterior distributions of $R(\mu_{tj})$ are approximated by
computing $\sum_{j' \in O_j}I[\mu_{tj'} \le\mu_{tj}]$
from samples of $[\mu_{t1}, \ldots,\mu_{tJ}]$ drawn from the multivariate
Gaussian with mean $[\hat{\mu}_{t1}, \ldots,\hat{\mu}_{tJ}]$ and variance
$\hat{\Sigma}_{\mu}$.
We define the posterior probability of a rank increasing ($U$) or
decreasing ($D$) DSE as $\operatorname{Pr}(R(\mu_{t_1 j})>R(\mu_{t_2 j}))=D_{t_1t_2 j}$
and $\operatorname{Pr}(R(\mu_{t_1j})<R(\mu_{t_2j}))={U}_{t_1t_2j}$ and estimate
these with the Monte-Carlo sample proportions satisfying these events.
If $\max(U_{t_1 t_2j}, D_{t_1 t_2j}) >\kappa$ where
$\kappa$ is a cutoff (say 0.9), then we declare junction $j$
differentially spliced between
tissues $t_1$ and~$t_2$.\

\begin{figure}

\includegraphics{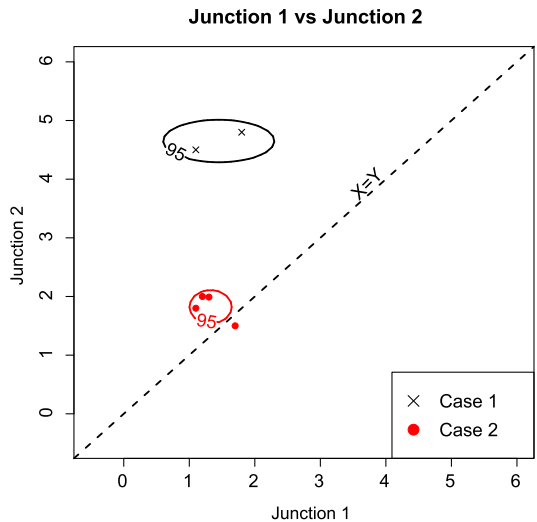}

\caption{A hypothetical experiment demonstrating that
the posterior ranks of the means provide more evidence compared to the
ranks of the raw observations.
The $95\%$ HPDs for the posterior means are drawn as ellipses.
In both cases, the prevalence of Junction 2 is greater than Junction 1.
In case 1, there are only 2 observations, but the posterior
distribution strongly favors
higher rank for Junction 1. However, this frequency of observed ranks
would randomly occur $25\%$ of the time.
In case 2, there is $>95\%$ posterior probability that Junction 2 has
higher mean than Junction 1,
even though the observed relative frequency that Junction 2 is greater
than Junction 1 is only $75\%$.
The accuracy of the evidence from the posterior depends upon parametric assumptions.}
\label{fig:rankpost}
\end{figure}

We use the term \textit{latent ranks} because it is important to
distinguish the proposed method of ranking the posterior means from the
typical use of rank statistics
computed directly from the observed values.
For example, \citet{geman2005} use rank reversals for microarray
classification in a method called Top Scoring Pairs, and they
compute the ranks of the raw microarray intensities and characterize
the variability of the ranks based upon the observed frequencies of orderings.
Given that microarray experiments have substantial measurement error as
well as potentially large effect sizes, we need to maximize the
accuracy and carefully quantify variability in rank estimates. When the
sample sizes are small as is often the case in differential splicing
experiments,
using the ranks of the posterior means may have advantages over using
the frequencies of the orderings. Figure \ref{fig:rankpost}
shows two scenarios of how the ranks of posterior means lead to
different results than using the observed ranks. The axes are the log expression
levels of 2 junctions, and the $95\%$ highest posterior density
(HPD) regions of the posterior means are outlined. Case~1 demonstrates
that one may estimate the posterior rank of the mean
of Junction~2 to be higher than Junction~1 with $95\%$ confidence given
only two observations if Junction~2 is much greater than 1 relative to
the error variance.
However, if one considered relative
frequencies of the ranks of two randomly ordered pairs, then the
pattern would occur $25\%$ of the time. In case 2, there are four observations,
and~3 out of 4 show Junction 2 having higher rank than 1. Because the
posterior density has a small variance, the posterior mean of
Junction 2 has a $95\%$ probability of being higher than Junction 1
even though $25\%$ of the observations show this not to be the case.
RCD maintains high power in low sample size compared to other ranking methods
because it does not use the raw ranks as it takes into account the
magnitude of differences relative to the variances and not just the
observed ordering.
Therefore, using the posterior ranks of means gives different results
from using the raw ranks, and this is appropriate for \textit{detecting
associations} of
relative junction prevalence with biological condition in experiments
with small sample sizes.
However, using the frequency of observed ranks as in \citet{geman2005}
is more appropriate for \textit{discriminating } biological
states for the purposes of classification with relatively large sample sizes.

The accuracy of the posterior relies upon distributional assumptions
and the
reasonableness of the prior.
The distributional assumption of the normality of residuals is an
accepted approximation
for microarrays [\citet{Cline2005}; \citet{kerr01}], and we found this is
reasonable for our data.
We chose a noninformative prior based upon the MLE approximation,
but one may tune the influence of the prior by altering the threshold
$\kappa$ to be more or less conservative.
It is conceivable to avoid the use of priors by constructing a
frequentist test for latent rank, but
this could be relatively cumbersome given the difficulty of maximum
likelihood estimation on the
discrete rank space.

The model identifies the event that the prevalence of one junction
surpasses or becomes less than another junction in different tissues.
This model detects a shift in intensity ratios between junctions such
as 1:2 changing to 2:1, but the model does not detect changes in ratios such
as 3:5 changes to 4:5. This way, the rank change model tests \textit{
qualitative} changes in prevalence such as when isoforms go from
dominant (highest rank) to nondominant (lesser rank)
rather than detecting
merely \textit{quantitative} changes in junction prevalence. We suggest
that such qualitative changes have more biological
impact and are worth specific detection algorithms. The RCD method is
implemented in the R programming language, and is
available at \url{http://sites.google.com/site/gbiostats/}.

\section{Simulation study}

We performed a simulation to directly compare the performance of the
RCD method to ANOSVA in terms of the false positive rate.
We anticipated that ANOSVA would perform well in the absence of
nonlinearity, but
in the presence of nonlinearity, it would have an unacceptably high
false positive rate.
We simulated data under 4 different scenarios: 2 incompatible isoforms
and linear response, 2 isoforms and
nonlinear response, 3 incompatible isoforms and linear response, and 3
incompatible isoforms
and nonlinear response. We used the previously discussed ANOSVA model
for simulation
$E[y_{tjr} ] =\mu_{tj}= \alpha_{t} + \beta_{j} + \gamma_{t \times j}$.
In each case, we had 2 tissue types and 12 two color arrays in a
balanced, dye-swap design. The tumor tissue differential
expression effect ($\alpha_t$) was simulated to consist of a log
fold-change using a Gaussian distribution of mean 0, standard deviation~1,
and the normal tissue level was drawn from the empirical distributions
of the means of the normal cells in our data set.
The relative prevalences of the incompatible junctions ($\exp(\beta
_{j})$) were simulated from
different Dirichlet distributions with parameters (1,1) for 2
junctions or (1,1,1) for 3 junctions.
Because we are estimating the false positive rates,
the simulation examines scenarios such that there are no differential
splicing events ($ \gamma_{t \times j}=0$).
The random error was an empirical distribution of the residuals of the
reference channel with mean 0 and standard deviation of $0.22$, and
no spot specific effects were simulated, as these effects were observed
to be relatively small in the data.
The nonlinear effect was simulated by transforming $E[y_{tjr} ]$ with a
linear transformation of a logistic
function that approximates the dynamic range of the array $P(x) =
w/(1+\exp(-(x-\mu^*)/\sigma^*))+\delta_{\mathrm{min}}$
where $w=9.2$ is the width of the observed dynamic range on the log
scale; $\delta_{\mathrm{min}} = 6.3 $ is the minimum of the dynamic range;
$\mu^*=10.9$ is the midpoint between the minimum and maximum of the
dynamic range; and $\sigma^*$ is selected so that $P(x)$ has slope${}= 1$ when
$x=\mu^*$.
The estimate of the false positive rate was based upon a $p$-value cutoff
of 0.05 for the ANOSVA model ($\gamma_{t \times j}=0$),
and a cutoff of posterior probability of unequal ranks of $\kappa=
0.9$ for the rank change model. We obtained the false positive
rates for 1000 simulations as shown in Table \ref{Tab:03}.

The false positive rate of ANOSVA is
close to the nominal 0.05 value under the linear scenarios, albeit
somewhat inflated
due to slight violations of parametric assumptions, but ANOSVA has a
substantially inflated
false positive rate of 3--4 times higher than the nominal value under
the nonlinear scenarios. In contrast,
the cutoff of 0.9 probability of differential splicing has a tolerably
low false positive rate.
The simulation suggests that the nominal $p$-values obtained for
differential splicing detection
could lead to an unacceptably high rate of false positive calls in the
presence of nonlinearity.
In general, we found that estimation of larger numbers of mutually
incompatible junctions resulted in more false
positives for both models. For this reason, we recommend that the
maximum size of $O_j$ considered be less than 10
or to adjust the posterior probability cutoff for larger $O_j$.

\begin{table}
\caption{False positve rate for RCD model and conventional ANOSVA
model}\label{Tab:03}
\begin{tabular*}{250pt}{@{\extracolsep{\fill}}lcc@{}}
\hline
 & \multicolumn{2}{c@{}}{\textbf{Model}}\\[-6pt]
& \multicolumn{2}{c@{}}{\hrulefill}\\
\textbf{Scenario}&\textbf{ANOSVA} & \textbf{RCD} \\
\hline
2 junction linear&0.063&0.000\\
2 junction nonlinear &0.165&0.008\\
3 junction linear&0.057&0.001\\
3 junction nonlinear&0.194&0.010\\
\hline
\end{tabular*}
\end{table}

We also performed a power analysis to demonstrate that the power loss
due to
the use of ranks is not too great compared to ANOSVA.
We considered power in both the linear and the nonlinear cases for two
opposing junctions, and we
simulated random error and tissue effects based upon empirical values
as previously described.
The interaction between tissue and junction was selected to be rank
reversals of two opposing junctions
such that a ratio of prevalence of $x\dvtx y$ in one tissue is reversed to
$y\dvtx x$ in the other tissue.
Without loss of generality, if $x=1$, then the effect size of the
interaction log scale would be $2\log_2(y)$.
We varied the effect size from $\log_2(1)$ to $2\log_2(8)$ for the
nonlinear case and from
$\log_2(1)$ to $2\log_2(1.5)$ in the linear case, and varied the sample
size from $n=4$ to $12$.
When the interaction effect size is $0=\log_2(1)$, the junction effect
is equal in both tissues and chosen at random
from the range of $\log_2(1.5)$ to $\log_2(8)$ for the nonlinear case
and from
$\log_2(1.05)$ to $\log_2(1.5)$ in the linear case.
The results of $1000$ simulations per scenario are shown in
Supplementary Figure 2. In the nonlinear case, ANOSVA has
Type I error rate of $0.212$ and $0.308$ for sample sizes of 4 and 12,
respectively, compared
to $0.004$ and $0.000$ for the RCD model. The power comparison in the
nonlinear case favors ANOSVA by
approximately $10\%$, although this is balanced by an increase in the
Type I error of more than 10-fold.
In the linear scenario, the relative power of ANOSVA vs. RCD is similar
to the nonlinear scenario while the sensitivity of both methods to
detect DSE is increased.
In the power comparison, there are two issues that need to be
emphasized. First,
the RCD model detects only cases when there is a qualitative difference
in rank so that
quantitative changes in prevalence such as 1:2 to 1:3 cannot be
reliably detected. Second,
the posterior probabilities given by RCD and the $p$-values of ANOSVA
have different meanings,
and this complicates direct comparisons.

\section{Application to glioblastoma data}

We studied one normal brain cell line (FGG) and four glioma lines
(U87MG, U118MG, T98G and A172).
The arrays used in our analysis were designed by JIVAN Biologicals (San
Francisco, CA) and manufactured by Agilent (Santa Clara, CA).
They contain 2145 genes identified to be related to cancer by
literature search. There were 38,425 splice junctions or potential
alternative splicing
events represented on the arrays as probes of length 35--40~bp. The probe
sequences were based upon build hg17 and can be downloaded with the raw
and normalized data from the Gene Expression Omnibus (GPL10127 and GSE20723).
200~ng of each RNA sample were labeled with Cy3/Cy5 dye using Agilent
Low RNA Input Linear
Amplification kit as per manufacturers protocols. After
amplification, 750~ng each of Cy3 and Cy5 labeled RNA were combined and
hybridized to microarrays.
A two color microarray based gene expression analysis protocol
(Agilent) was used.
The normal FGG line was used as a reference
for all arrays in a dye-swap design with the 4 glioma lines being
labeled Cy3 and Cy5 two times each for a total of 16 arrays.
See Supplemental Table~1 for a layout of the experimental design.
Microarrays were scanned using the Agilent Microarray Scanner G2565AA
and extracted using Agilent feature Extraction Software version 9.1.
Array data preprocessing and normalization were performed using the
SpliceFold software JIVAN Biologicals (San Francisco, CA).

\begin{figure}

\includegraphics{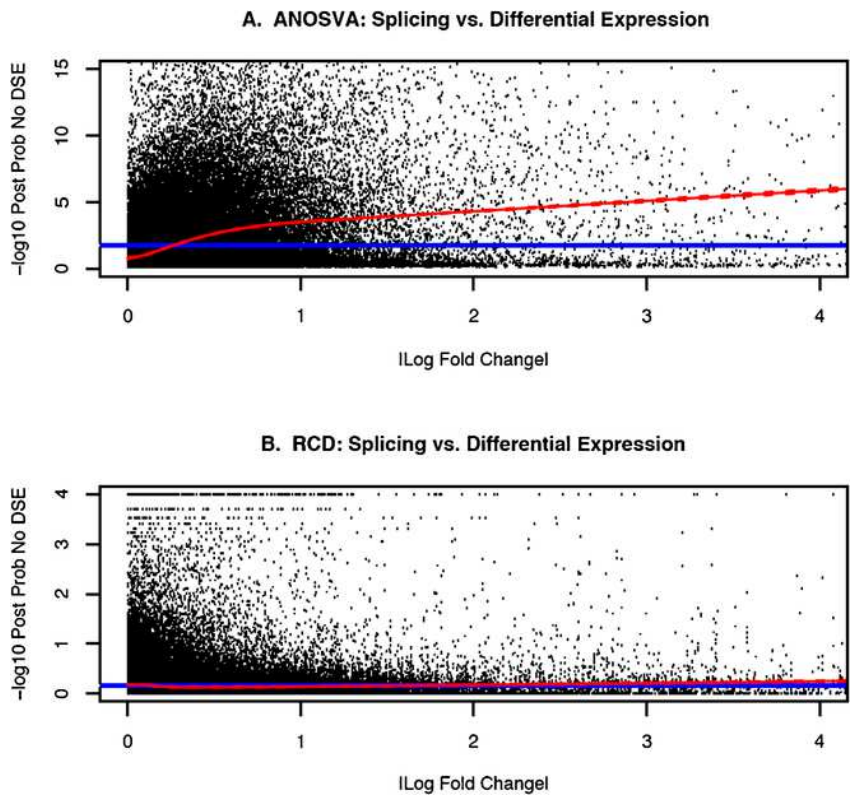}

\caption{Splicing Probability vs differential expression Log Fold Change.
(\textup{A}) The ANOSVA model estimate of local false discovery rate for
differential splicing ($-\log_{10}$ Posterior Probability of No DSE)
has a strong dependency on differential expression fold change
($|\log\mbox{Fold Change}|$). A spline curve is shown in red, and
the mean $-\log_{10}$ lFDR
of differential splicing is in blue.
This indicates that the higher fold change for differential expression
implies higher probability of differential
splicing. The nonlinear effects within the microarray confound the hypotheses
about splicing and differential expression.
(\textup{B}) The RCD model estimate for differential splicing probability
has little dependency on overall changes in expression.
A spline curve is shown in red, and the mean $-\log_{10}$ posterior probability
of no differential splicing is in blue.}\label{fig:ASvDE}
\end{figure}

The size of the sets of incompatible junctions considered ranged from 2
to 10. The percentages of sets with 2, 3, 4 and 5 junctions
were 31\%, 19\%, 11\% and 8\%, respectively. We examined the
incompatible sets corresponding to each junction.
The RCD model fit indicates that hundreds of junctions have posterior
probability for rank change of $>0.9$; see Table \ref{Tab:02}A.
Note that some events counted may be interrelated, as some sets of
incompatible junctions have nonnull intersections.
A~full list of the significant genes and junctions are downloadable
from the url.
We also compared the results with the ANOSVA method shown in Table \ref{Tab:02}B.
Here, a junction is called differentially spliced if the
false discovery rate (FDR) estimated by the qvalue is
less than 0.1 [\citet{Stor03}]. The qvalue criterion is inherently more
liberal than the posterior probability criterion, but this
difference is modest.
Table \ref{Tab:02}C demonstrates the overlapping significant junctions satisfying
both criteria for the RCD and the ANOSVA models. Note that most of the
junctions identified as significant according
to the rank change model are declared significant according to ANOSVA model.
The number of junctions identified by ANOSVA is much higher, although,
we argue, that
the results are heavily confounded with the substantial differential
expression within the experiment.
To demonstrate the confounding between differential expression and the
differential splicing events
estimated by ANOSVA, we tested the hypothesis that the junctions that
were differentially expressed were more likely to be declared a DSE.
By differentially expressed, we mean that the average expression in one
cell line is higher than the other for a given set of junctions.
Figure \ref{fig:ASvDE} shows evidence for differential splicing events
estimated by ANOSVA and the RCD
model and their relationship
to the estimated differential expression.
The evidence for a DSE for the ANOSVA model is quantified by the local
false discovery rate (lFDR) [\citet{pounds2006}].
In Figure \ref{fig:ASvDE}A, the ANOSVA model estimate of local false
discovery rate for
differential splicing (Posterior Probability of No DSE) has a strong
dependency on the
log fold change estimate for differential expression. This indicates
the higher fold change estimates for
differential expression imply higher posterior probability of differential
splicing. This is consistent with the idea that the nonlinear effects
within the microarray confound the hypotheses
about splicing and differential expression, and that the estimates of
differential splicing are biased and
substantially overestimated.
In Figure \ref{fig:ASvDE}B, the RCD model estimate for differential
splicing probability
has little dependency on overall changes in expression. This is
consistent with the robustness of the
rank change model to bias due to change in expression level.

\begin{table}
\caption{Number of alternative splicing events between cell types for
RCD and ANOSVA models}\label{Tab:02}
\begin{tabular*}{\textwidth}{@{\extracolsep{\fill}}lccccc@{}}
\hline
&\textbf{Cell type}&\textbf{FGG} & \textbf{A172} & \textbf{T98G} & \multicolumn{1}{c@{}}{\textbf{U118}} \\
\hline
A. RCD&A172 &\phantom{0}364&&\\
&T98G &\phantom{0}344&\phantom{00.}365\\
&U118 &\phantom{0}389&\phantom{00.}348&\phantom{0}399&\\
&U87 &\phantom{0}333&\phantom{00.}334&\phantom{0}311&\phantom{0}264\\[3pt]
B. ANOSVA&A172 &6058&&\\
&T98G &7610&11,106\\
&U118 &7771&\phantom{0.}8695&8890&\\
&U87 &6590&\phantom{0.}8319&7546&5048\\[3pt]
C. RCD \& ANOSVA&A172 &\phantom{0}267&&\\
&T98G &\phantom{0}286&\phantom{00.}346\\
&U118 &\phantom{0}351&\phantom{00.}316&\phantom{0}367&\\
&U87 &\phantom{0}301&\phantom{00.}282&\phantom{0}274&\phantom{0}213\\
\hline
\end{tabular*}
\end{table}

We also examined whether the genes with splicing events reported in the
literature for glioblastoma had
greater posterior probability for splicing under the ANOSVA and the RCD
models. \citet{Cheung2009} reported
such genes from multiple studies, including 8 that were assessed by our
arrays: CALD1, CASP2, FGFR1, NF1,
RAB3A, ST18, TNC and TPD52L2. For all comparisons between cell types,
we computed the
\textit{enrichment ratio} which we define as the proportion of
significant DSEs in the previously reported genes relative to
proportion of significant DSEs in all other genes at various cutoffs.
The cutoffs we used for the RCD model were posterior probabilities for
rank change greater than 90\%, 99\%, 99.9\% and 99.99\%,
and the cutoffs for ANOSVA were lFDR values of less than $10^{-1,-2,-3,
\ldots,-15}$. The results are in Supplemental Tables 2 and 3.
The enrichment ratio peaked at 3.17 for ANOSVA when the cutoff was
$10^{-14}$, giving 36 significant DSEs in
the known genes. The enrichment ratio plateaued at 13 for RCD when the
cutoff was 99.99\%, giving 18 significant DSEs in
the same genes. However, at the corresponding cutoffs in RCD and ANOSVA
resulting in 36 significant DSEs in the known genes,
the RCD enrichment ratio was 6.7 compared to 3.17 for ANOSVA. This
implies that
the RCD method found that the known genes were substantially more
enriched for DSEs than the ANOSVA method
for equally conservative cutoffs. We repeated the above enrichment
score estimation with
randomly selected sets of 8 genes, and found that $<1\%$
of these sets give a DSE enrichment score greater than the known
genes.\looseness=1

\section{Discussion}

We presented a model for the detection of differential splicing events
in the presence of substantial nonlinear
effects of the microarray intensity response. This model is robust to
the violations of the assumption of linearity
and is designed to detect features of alternative splicing that are
invariant under monotonic transformation.
One difficulty with the model is that differential splicing may occur
without exhibiting the invariant feature of
rank change. Although these features are likely present in the most dramatic
and biologically important differential splicing events, the proposed
model may have lower power to detect
more subtle shifts in the proportions of isoforms expressed in
different tissues. We choose to trade sensitivity
for specificity in the current setting as the false positive rate
(though not knowable) is potentially quite high
for other methods. This method may be used with either one- or
two-color splice junction array technology
and can be applied to data with or without nonlinear effects to
identify \textit{qualitative} changes in junction
prevalence.
It may be argued that filtering out either very low or very high
measurements from downstream
analysis would reduce the degree of nonlinearity in the data. However,
the selection of a cutoff would be
difficult and error prone. Furthermore, very high and very low
intensities can be informative so that
the removal of meaningful measurements degrades the signal in the data.

There are several opportunities to extend the proposed model.
We estimated the posterior
distribution based upon the MLE, and this has the effect of selecting
noninformative priors.
Noninformative priors imply that there is an equal probability of there
being an alternative splicing
event or not, but the accuracy of the method could be improved if one
utilized data on the prior probability
of particular alternative splicing events.
The hypothesis of rank change
can be extended to any model for which there are posterior
distributions of parameters related to the
ranks of relative isoform prevalence.
First, we may explicitly model the sigmoidal
response of the microarray intensity with some appropriate parametric
link function $g(\cdot)$ so that we have
$g(E[y_{tjr} ]) = \mu_0 + \alpha_{t} + \beta_{j} + \gamma_{t \times
j}$. This would likely add to the computational challenge of fitting
the model, but it would improve sensitivity to
detect DSEs.
If we next consider $g(\cdot)$ as a link function in a log-linear count model,
we may model splice junction counts from RNA-seq data instead of
microarray intensity.
Second, we had assumed that the error terms $\varepsilon_{tjr}$ were
independent of one another,
although it is likely that the incompatible isoforms have errors that
are negatively correlated, and these
correlations of the residuals can be modeled using a compound symmetric
structure.
Third, we may extend the model to include knowledge of the transcripts
by adding a latent variable
for each known isoform such that
%
\begin{equation}
y_{tjr} = \sum_{h}\beta_{j} z_{ht} \delta_{hj} + \nu_{tjr} +\varepsilon_{tjr},
\end{equation}
where $\beta_j$ is the effect of junction $j$, $z_{ht}$ is the latent
variable proportional to
the amount of isoform $h$ within the tissue $t$, and $\delta_{hj}$ is
the indicator
variable for whether or not junction $t$ is within junction $h$. This
extension could incorporate
exon probes as well. The latent variable $z_{ht}$
would be an indication of the prevalence of the isoform, and estimating
the ranks and
differences in $z_{ht}$ would help to identify differential splicing
events by pooling
information across junctions and exons within a complete isoform.

\section*{Acknowledgments}

We thank the editors and two anonymous referees for suggesting many
improvements to the paper.
Many thanks to Jonathan Bingham at JIVAN Inc. for helpful discussions
and insights
into this novel array platform. The Greehey Children's Cancer Research Institute
and the Cancer Therapy \& Research Center provided laboratory and other
resources to LOFP.
This work (JG) was supported by CTSA Award Number [KL2 RR025766] from
the National Center for Research Resources.
The content is solely the responsibility of the authors and does not
necessarily represent the official views of the National Center for
Research Resources of the National Institutes of Health.

\begin{supplement}[id=suppA]
\stitle{Supplemental Figures and Tables}
\slink[doi]{10.1214/10-AOAS389SUPP}
\slink[url]{http://lib.stat.cmu.edu/aoas/389/supplement.pdf}
\sdatatype{pdf}
\sdescription{Supplemental Table 1. Experimental layout that describes the number of samples, arrays, and dye
orientation.
Supplemental Table 2. Enrichment ratio describing the proportion of known differential splicing events
detected by ANOSVA.
Supplemental Table 3. Enrichment ratio describing the proportion of known differential splicing events
detected by RCD.
Supplemental Figure 1. Graphic describing the common forms of alternative splicing.
Supplemental Figure 2. Power comparison of ANOSVA and RCD for different effect sizes, sample sizes,
and degree of nonlinearity.}
\end{supplement}

%
%

\printaddresses

\end{document}